\documentclass[epj,nopacs]{svjour}
\usepackage{graphics}
\begin{document}
\title{Jet quenching pattern at LHC in PYQUEN model}
\author{I.P. Lokhtin, A.V. Belyaev \and A.M.Snigirev 
}                     
%
%
\institute{D.V. Skobeltsyn Institute of Nuclear Physics, M.V. Lomonosov Moscow State University, Moscow, Russia}
%
%
\abstract{
The first LHC data on high transverse momentum hadron and dijet spectra in PbPb collisions at 
center-of-mass energy 2.76 TeV per nucleon pair are analyzed in the frameworks of PYQUEN jet 
quenching model. The presented studies for the nuclear modification factor of high-p$_T$ hadrons and 
the imbalance in dijet transverse energy support the supposition that the intensive wide-angular (``out-of-cone'') 
medium-induced partonic energy loss is seen in central PbPb collisions at the LHC. 
%
} 
%
\maketitle
\section{Introduction}

The production of high transverse momentum hadrons and jets is one of the important 
probes of hot quark-gluon matter (QGM) created in ultrarelativistic heavy ion collisions. 
The energy loss of energetic partons, so called ``jet quenching'', 
is predicted to be very different in cold nuclear matter and in hot QGM, 
and leads to a number of phenomena which are already seen in gold-gold collisions at 
RHIC ($\sqrt s_{\rm NN}=200$ GeV), such as 
suppression of high-$p_T$ hadron spectrum, medium-modified angular correlations, azimuthal 
anisotropy, etc. (see, e.g., recent reviews \cite{d'Enterria:2009am,Wiedemann:2009sh,Accardi:2009qv,Majumder:2010qh,Dremin:2010jx} 
and references therein). The current lead-lead collision energy at LHC, $\sqrt s_{\rm NN}=2.76$ TeV, 
is a factor of $\sim 14$ larger than that in RHIC, thereby allows one to probe new frontiers of 
super-high temperature and (almost) net-baryon free Quantum Chromodynamics (QCD). It is expected 
that at such ultra-high energies the role of hard and semi-hard particle production may be 
significant even for the bulk properties of the created matter. 

The first striking results of LHC heavy ion program are the momentum-dependent suppression of charge 
high-p$_T$ particle production~\cite{Aamodt:2010jd} and the large imbalance in 
dijet transverse energy~\cite{Aad:2010bu,Chatrchyan:2011sx}. It was supposed 
in~\cite{CasalderreySolana:2010eh} with simple kinematic arguments that a frequency 
collimation mechanism in medium-modified parton showering can account for these 
observables. The calculation of dijet asymmetry due to multiple scatterings of shower partons to large angles has been done in~\cite{Qin:2010mn}. The dijet asymmetry was also computed recently with MARTINI model in~\cite{Young:2011qx}. The decoherence of parton showering in a dense medium was considered in~\cite{MehtarTani:2011tz}. Some numerical predictions for jet broadening in PbPb collisions at the LHC can be found in~\cite{Vitev:2008rz}. In present paper, the jet quenching pattern at the LHC is analyzed in the frameworks of PYQUEN partonic energy loss model~\cite{Lokhtin:2005px}, and the related interpretation 
of the data is discussed. 
 
\section{PYQUEN partonic energy loss model}

A number of Monte-Carlo models to combine a perturbative final state parton shower with QCD-medium effects 
and simulate the jet quenching in various approaches have been developed in recent years~\cite{Lokhtin:2005px,Zapp:2008gi,Zapp:2008af,Armesto:2008qh,Schenke:2009gb,Renk:2010zx}. Our model PYQUEN 
(PYthia QUENched)~\cite{Lokhtin:2005px} was constructed as a modification of the jet event obtained with the 
generator of hadron-hadron interactions PYTHIA$\_$6.4~\cite{Sjostrand:2006za}. The details of PYQUEN physics model 
and simulation procedure can be found in~\cite{Lokhtin:2005px}, the main features of the model being listed 
only very briefly below.

The approach to the description of medium-induced multiple scattering of hard partons in PYQUEN 
is based on the accumulated energy loss via the gluon radiation being associated with each parton 
scattering in the expanding quark-gluon fluid and includes the interference effect (for the emission 
of gluons with a finite formation time) using the modified radiation spectrum $dE/dx$ as a function 
of decreasing temperature $T$. The model takes into account the radiative and collisional energy loss of hard 
partons in longitudinally expanding quark-gluon fluid, as well as a realistic nuclear geometry. 
The radiative energy loss is treated in the frameworks of BDMS 
model~\cite{Baier:1999ds,Baier:2001qw} with the simple generalization to a massive quark case 
using the ``dead-cone'' approximation~\cite{Dokshitzer:2001zm}. The collisional energy loss due to elastic 
scatterings is calculated in the high-momentum transfer limit~\cite{Bjorken:1982tu,Braaten:1991jj,Lokhtin:2000wm}. 
The strength of the energy loss in PYQUEN is determined mainly by the initial maximal temperature $T_0^{\rm max}$ of hot
matter in central PbPb collisions (i.e. an initial temperature in the center of nuclear overlapping area at mid-rapidity). 
The energy loss depends also on the proper time $\tau_0$ of QGM formation and the number $N_f$ of active flavors in the 
medium. The parameter values $\tau_0=0.1$ fm$/c$ and $N_f=0$ (gluon-dominated plasma) are used for the simulations presented.  

Another important ingredient of the model is the angular spectrum of medium-induced radiation. Since the 
detailed treatment of this is rather sophisticated, the simple parametrizations of the gluon distribution over the emission angle $\theta$ are used. There are two main 
options in PYQUEN. The first one is the ``small-angular'' radiation,
\begin{equation} 
\label{sar} 
\frac{dN^g}{d\theta}\propto \sin{\theta} \exp{\left( 
-\frac{(\theta-\theta_0)^2}{2\theta_0^2}\right) }~, 
\end{equation} 
where $\theta_0 \sim 5^0$ is the typical angle of the coherent gluon radiation as estimated 
in~\cite{Lokhtin:1998ya}. The second one is the ``wide-angular'' radiation,  
\begin{equation} 
\label{war} 
\frac{dN^g}{d\theta}\propto 1/\theta~, 
\end{equation} 
which is similar to the angular spectrum of parton showering in a vacuum without coherent 
effects~\cite{Dokshitzer:1991wu}. 
In this paper, in order to illustrate the sensitivity of experimental observables to various jet quenching scenarios, the third (rather extreme) option ``extra wide-angular'' radiation, 
\begin{equation} 
\label{x-war} 
\frac{dN^g}{d\theta}\propto 1/\sqrt{\theta}~, 
\end{equation} 
is also considered. One can assume that the physical meaning of (extra) wide-angular radiation could be the 
presence of intensive secondary rescatterings of in-medium emitted gluons~\cite{Qin:2010mn}.
The collisional energy loss is considered in PYQUEN as ``out-of-cone'' loss 
(i.e. an energy is ``absorbed'' by the medium, because the major part of  ``thermal'' particles knocked out of the 
dense matter by elastic rescatterings fly outside a typical jet cone~\cite{Lokhtin:1998ya}). 

The event-by-event simulation procedure in PYQUEN includes the following steps. 
\begin{itemize}
\item The generation of initial parton spectra with PYTHIA and production vertexes at 
the given impact parameter. 
\item The rescattering-by-rescattering simulation of the parton path in a dense zone 
and its radiative and collisional energy loss. 
\item The final hadronization according to the Lund string model for hard partons and 
in-medium emitted gluons. 
\end{itemize}

In presented studies, the effect of nuclear shadowing on parton distribution functions, calculating the 
nuclear modification factor $R_{\rm AA}$ for charged particles at $p_T > 10$ GeV, is 
taken into account. The procedure is similar to one applied in HYDJET and HYDJET++ 
event generators~\cite{Lokhtin:2008xi}. It uses the impact parameter 
dependent parametrization obtained in the frameworks of Glauber-Gribov 
theory~\cite{Tywoniuk:2007xy}. The nuclear shadowing correction factor 
$S(r_1, r_2, x_1, x_2, Q^2)$ depends on the kinematic variables of incoming hard 
partons (the momentum fractions $x_{1,2}$ of the initial partons from the incoming nuclei 
and the momentum scale $Q^2$ of the hard scattering), and on the transverse coordinates 
$r_{1,2}$ of the partons in their respective nuclei. This initial state effect reduces 
the number of partons in the incoming hadronic wave-function of both the nuclei 
and thus reduces the yield of high-p$_T$ hadrons from jet production (in addition to 
the medium-induced partonic energy loss). 

\section{Numerical results}

PYQUEN partonic energy loss model was applied to calculate the nuclear modification factor of 
charged hadrons and the imbalance in dijet transverse energy in central PbPb collisions at 
$\sqrt s_{\rm NN}=2.76$ TeV for various scenarios of medium-induced radiation. The numerical 
results were compared with the LHC data applying the same kinematic cuts as in the experiments.

\subsection{Nuclear modification factor}

\begin{figure*}
\resizebox{0.5\textwidth}{!}{%
\includegraphics{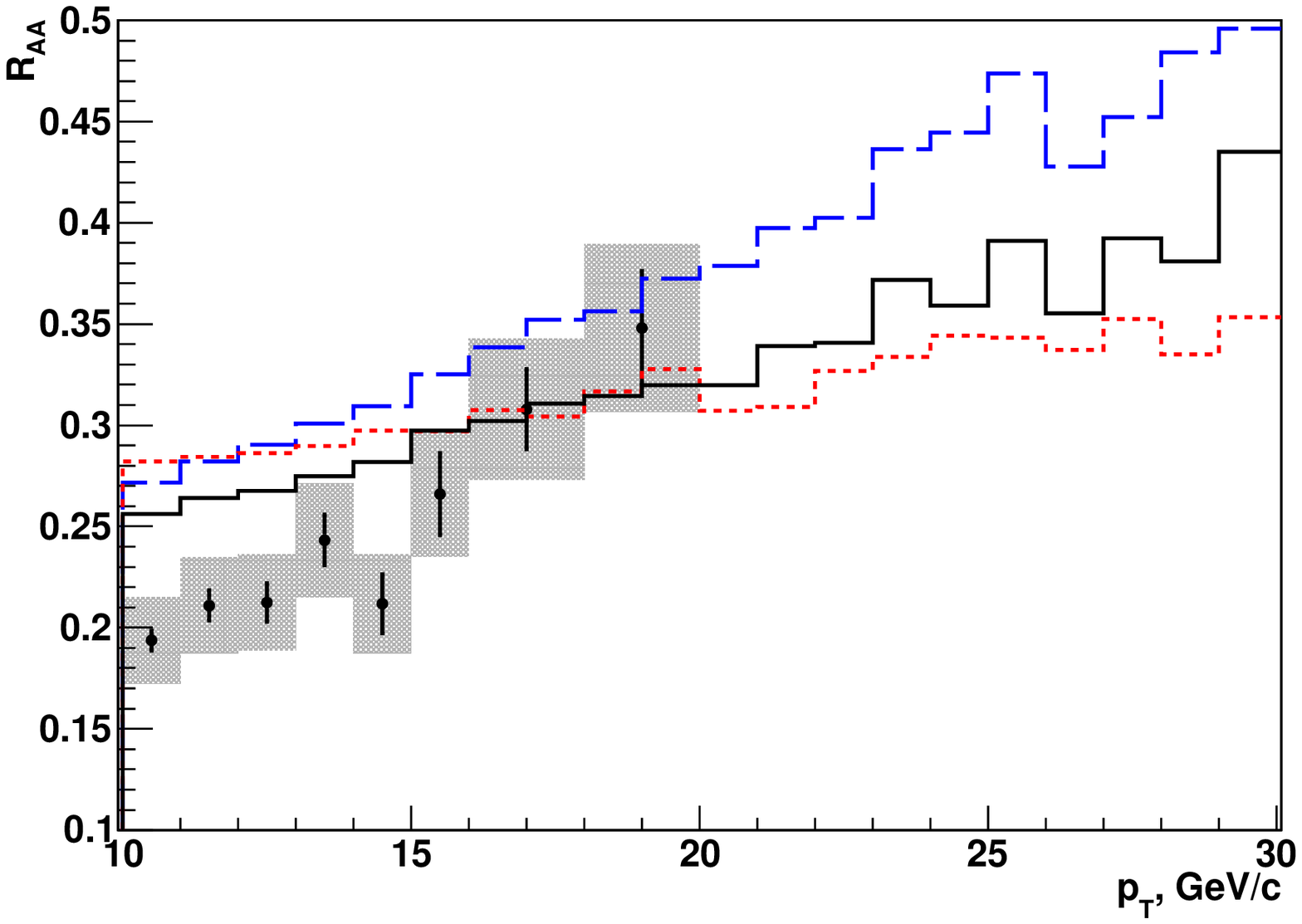}
}
\resizebox{0.5\textwidth}{!}{%
\includegraphics{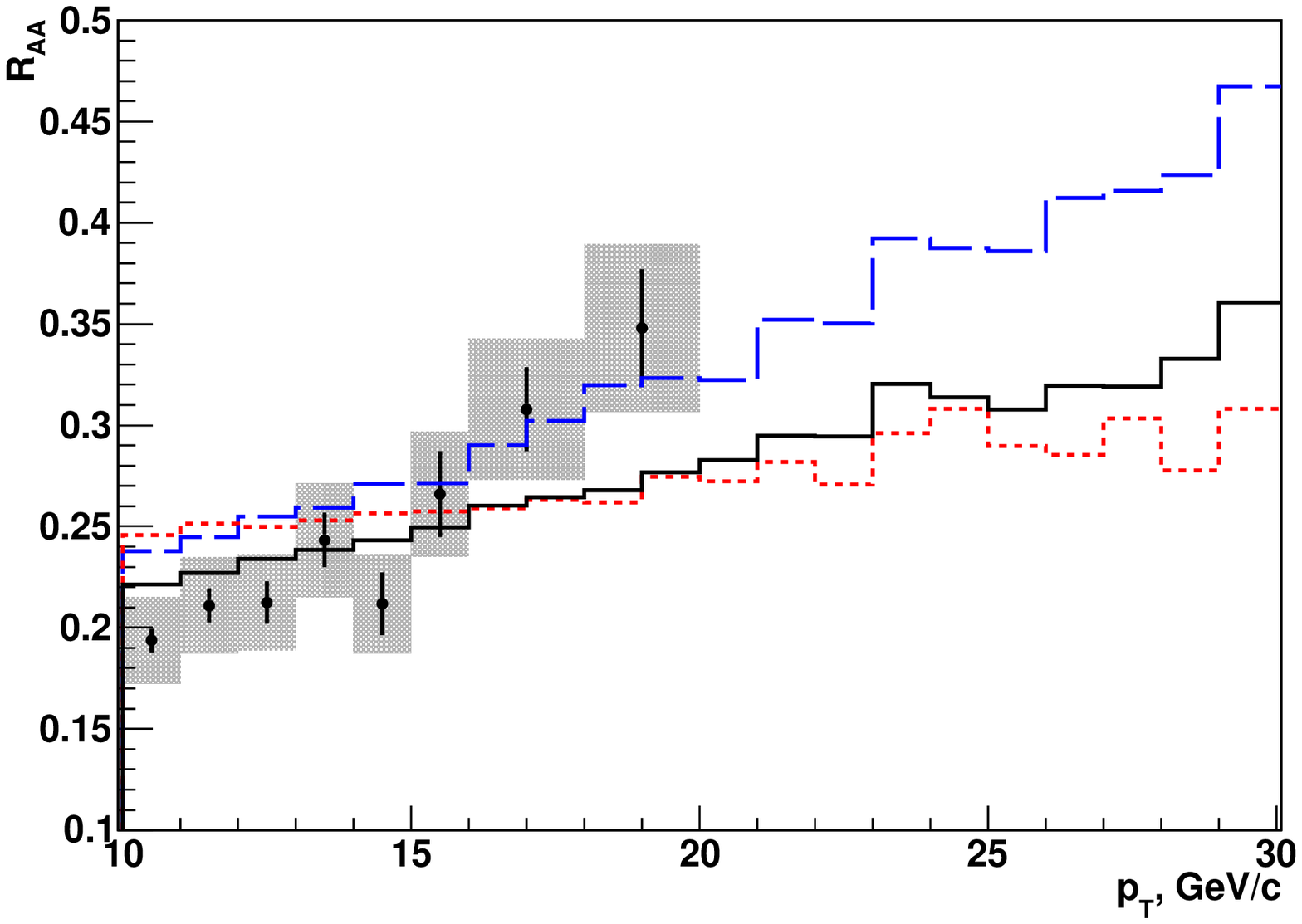}
}
\caption{The nuclear modification factor $R_{\rm AA}$ for charged hadrons in 5\% of most central PbPb collisions at $\sqrt s_{\rm NN}=2.76$ TeV. The points are ALICE data~\cite{Aamodt:2010jd} (the error bars show the statistical uncertainties, the boxes show the systematic errors), the histograms are simulated PYQUEN events. The dotted, solid and dashed histograms correspond to the parametrizations of the radiation angular spectrum 
(\ref{sar}), (\ref{war}) and (\ref{x-war}) respectively. The initial maximal temperature of quark-gluon matter $T_0^{\rm max}$=0.85 GeV (left) and 1 GeV (right).}
\label{raa-fig}      
\end{figure*}

The nuclear modification factor $R_{\rm AA}$ is defined as a ratio of particle yields 
in AA and pp collisions normalized on the number of binary nucleon-nucleon sub-collisions 
$\left<N_{\rm coll}\right>$: 
\begin{equation} 
\label{raa} 
R_{\rm AA}(p_T)=\frac{d^2N^{\rm AA}/d\eta dp_T}{\left<N_{\rm coll}\right>d^2N^{\rm AA}/d\eta dp_T}~,
\end{equation} 
where $p_T$ is the transverse momentum and $\eta$ is the pseudorapidity. In the absence of nuclear effects (in initial or final states) for high 
$p_T$ it should be $R_{\rm AA}=1$. Figure~\ref{raa-fig} shows the calculated nuclear modification 
factor $R_{\rm AA}$ for charged hadrons ($|\eta|<0.8$) as compared with ALICE data~\cite{Aamodt:2010jd} in 
5\% of most central PbPb collisions ($T_0^{\rm max}=$0.85 GeV for the left plot and 1 GeV for the right plot). Three options for the angular spectrum of in-medium gluon radiation (\ref{sar}), (\ref{war}) and 
(\ref{x-war}) were considered, collisional energy loss and nuclear shadowing being also taken into account. Higher initial temperatures result in more suppression of high-p$_T$ hadrons, while a wider angular radiation results in a 
stronger dependence of $R_{\rm AA}$ on $p_T$. The PYQUEN scenarios with $T_0^{\rm max}=$1 GeV 
and wide-angular radiation (\ref{war}) or (\ref{x-war}) seem closer to the data than other options. However 
even in this case, the simulated $p_T$-dependence of $R_{\rm AA}$ looks weaker than the tendency of the ALICE data. The large systematic uncertainties of the data (due to the absence of pp data at the same energy) still complicate a rigid quantitative interpretation of this observation. But if such discrepancy would be confirmed in the future with higher statistics and reference pp data at $\sqrt{s}=2.76$ TeV, in our opinion it could indicate that the real energy dependence of partonic energy loss $\Delta E$ is weaker than the expected one. It is worth to remind here that the energy dependence of BDMS radiative loss~\cite{Baier:1999ds,Baier:2001qw} (the dominant mechanism of medium-induced energy loss in PYQUEN) is between $\Delta E \propto \ln {E}$ and $\Delta E \propto \sqrt {E}$ (depending on the energy of the radiated gluon and on the properties of the medium). 

\begin{figure*}
\resizebox{0.5\textwidth}{!}{%
\includegraphics{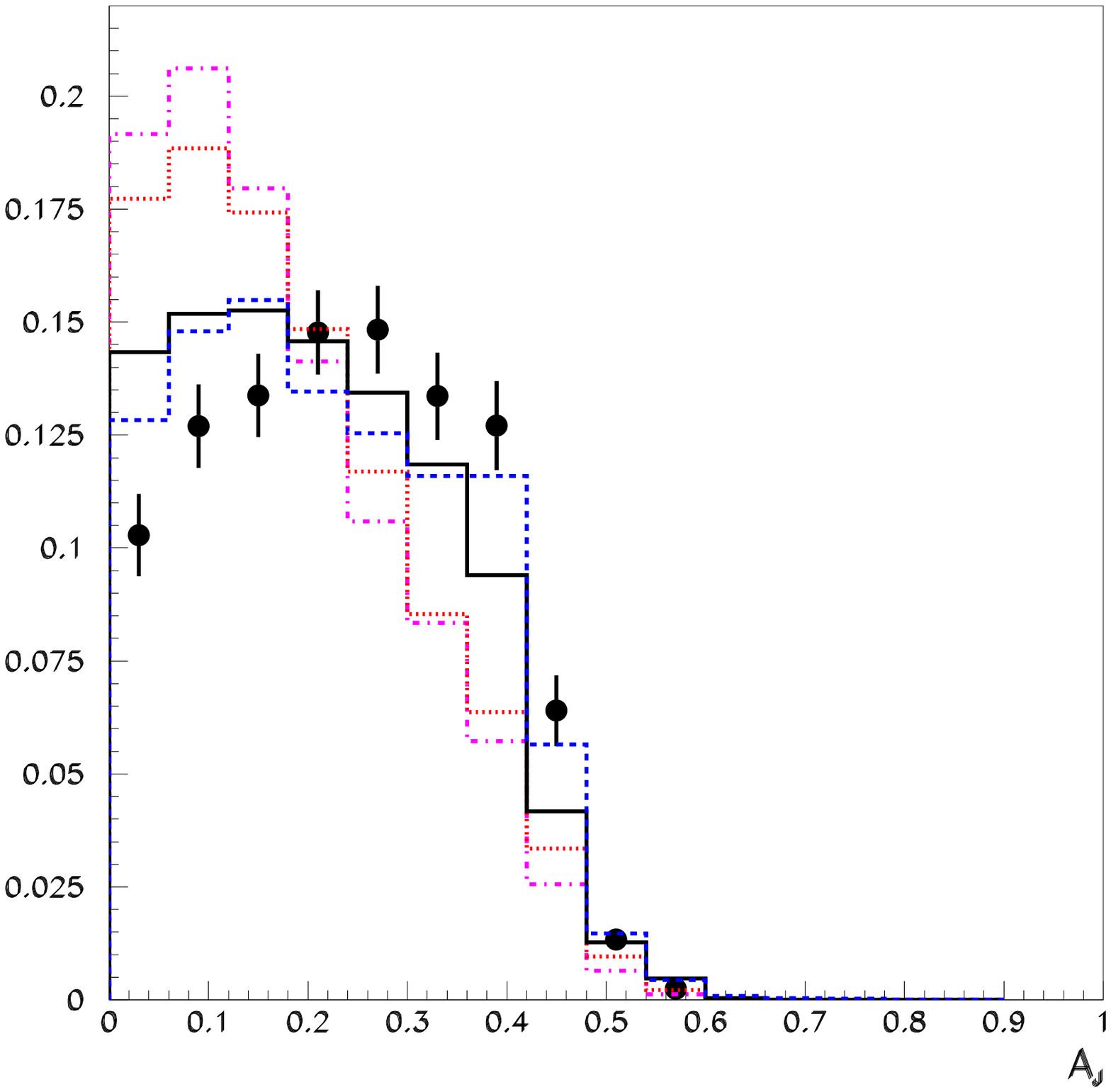}
}
\resizebox{0.5\textwidth}{!}{%
\includegraphics{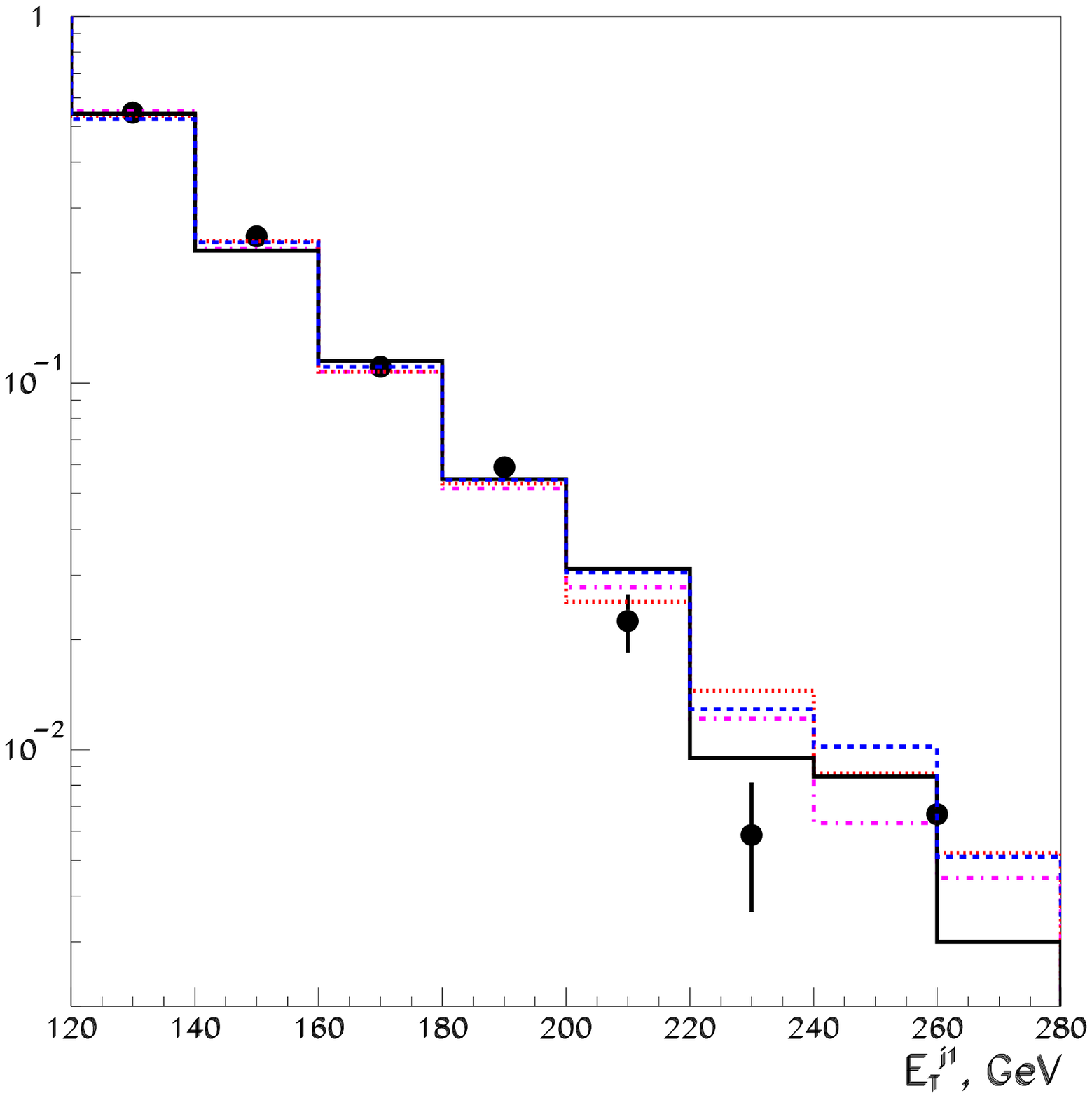}
}
\caption{The event fraction of dijet events with the asymmetry ratio $A_J$ (left) and with the leading jet 
energy $E_T^{j1}$ (right) in 10\% of most central PbPb collisions at $\sqrt s_{\rm NN}=2.76$ TeV. The points are CMS data~\cite{Chatrchyan:2011sx} (the error bars show the statistical uncertainties), the histograms are simulated PYTHIA (dash-dotted) 
and PYQUEN (other styles) events. The dotted, solid and dashed histograms correspond to the parametrizations of the radiation angular spectrum (\ref{sar}), (\ref{war}) and (\ref{x-war}) respectively. The initial maximal temperature of quark-gluon matter $T_0^{\rm max}$=1 GeV.}
\label{aj-fig}      
\end{figure*}

Let us discuss some other model-dependent uncertainties of the results obtained. Excluding the collisional loss from the simulations increases $R_{\rm AA}$ by $\sim 15\div 20$\% for $p_T=10\div20$ GeV/$c$. Excluding the nuclear shadowing increases $R_{\rm AA}$ by $\sim 30\div 20$\% for the same kinematic range and weakens its $p_T$-dependence. The choice of the medium formation time $\tau_0$ within its reasonable range has rather moderate influence on the strength of medium-induced partonic energy loss. We have found that increasing 
$\tau_0$ by factor $2$ (at fixed $T_0^{\rm max}=1$ GeV) reduces $R_{\rm AA}$ by $\sim 13$\% without changing its $p_T$-dependence. Note that the jet quenching effect gets stronger at larger $\tau_0$ due to slower cooling of the matter, which implies a jet parton spending more time in the hottest regions of the medium. As a result the rescattering intensity increases (it starts little later for larger $\tau_0$, but this factor is much  less significant). Fixing $\tau_0=0.2$ fm/$c$ and decreasing $T_0^{\rm max}$ to $0.85$ GeV allow us to get the similar results for $R_{\rm AA} (p_T)$ as shown in Figure~\ref{raa-fig} (right). 
The jet quenching effect is also sensitive to the space-time evolution of a dense matter. For calculations presented, the medium was treated as a boost-invariant longitudinally expanding perfect quark-gluon fluid (well-known scaling solution obtained originally by Bjorken~\cite{Bjorken:1982qr}, $T(\tau)=T_0(\tau_0/\tau)^{1/3}$). In principle, other scenarios of QGM space-time evolution for PYQUEN are also envisaged~\cite{Lokhtin:2005px,Lokhtin:2000wm}. For example, the presence of fluid viscosity slows down the cooling rate, i.e., in fact the effective temperature of the medium gets higher as compared with the perfect QGM case. On the other hand, the influence of the transverse expansion on the intensity of parton rescattering is inessential for high initial temperatures $T_0^{\rm max} \gg T_c$ ($T_c \sim 200$ MeV is the critical temperature). 

\subsection{Dijet transverse energy imbalance}

Dijet asymmetry ratio $A_J$ is defined~\cite{Aad:2010bu,Chatrchyan:2011sx} as a difference between 
transverse energies of two jets $j1$ and $j2$ normalized on its sum:
\begin{equation} 
\label{aj} 
A_J=\frac{E_T^{j1}-E_T^{j2}}{E_T^{j1}+E_T^{j2}}~.
\end{equation} 
In order to define calorimetric jets, we use the toy jet finding model. The jet energy was determined as the total transverse energy of the final particles collected around the direction of the leading particle inside the cone of radius 
$R=\sqrt{\Delta \eta ^2+\Delta \varphi ^2}=0.5$ ($\eta$ and $\varphi$ are the pseudorapidity and the 
azimuthal angle respectively). Then the final jet energy was smeared using Gaussian distribution with $\sigma$ 
proportional to the square root of a ``true'' jet energy. The same kinematic cuts as in the CMS analysis~\cite{Chatrchyan:2011sx} were applied: $|\eta^{j1,2}|<2$, $|\varphi^{j1}-\varphi^{j2}|>2.1$, $E_T^{j1}>120$ GeV, $E_T^{j2}>50$ GeV. Note that we compare our simulations with the CMS data having harder cuts on minimum 
jet energy than ATLAS data, in order to reduce complications facing jet reconstruction in high multiplicity background~\cite{Cacciari:2011tm,Young:2011qx}.  

Figure~\ref{aj-fig} (left) shows the calculated fraction of dijet events with the asymmetry ratio 
$A_J$ for three options of the angular spectrum of in-medium gluon radiation (\ref{sar}), (\ref{war}) and (\ref{x-war}) respectively. The choice of main parameters ($T_0^{\rm max}=$1 GeV, $\tau_0=0.1$ fm$/c$) was fixed from fitting the LHC data on $R_{\rm AA}(p_T)$ discussed in previous sub-section. The collisional energy loss is taken into account, the nuclear shadowing effect 
being negligible for such high E$_T$ jets. PYQUEN results are compared with CMS data~\cite{Chatrchyan:2011sx} for 10\% of 
most central PbPb collisions and with PYTHIA simulations (in fact, PYQUEN without medium effects). The option with the small-angular radiative energy loss (\ref{sar}) does not result in any additional dijet asymmetry as compared with PYTHIA (small 
additional asymmetry seen for dotted histogram is due to the collisional loss). Indeed, in this case the bulk of in-medium emitted gluons fly close to the parent parton direction and still belong to the jet. Therefore, the small-angular medium-induced radiation softens particle energy distributions inside the jet and increases the multiplicity of secondary particles, but does not affect the total jet energy~\cite{Lokhtin:1998ya}. On the other hand, the wide-angular radiation (\ref{war}) and (\ref{x-war}) generates the significant dijet asymmetry due to the substantial fraction of ``out-of-cone'' jet energy loss. We would not expect the perfect agreement of PYQUEN simulations with the data using our oversimplified treatment of jet finding (without taking into account realistic detector effects, jet reconstruction details and high multiplicity background fluctuations), but at least qualitative reproducing of the dijet asymmetry 
can be achieved with the wide-angular radiative and collisional partonic energy loss. This interpretation is in the 
agreement with the observation of CMS collaboration that the dijet momentum balance is recovered when integrating low transverse momentum particles distributed over a wide angular range relative to the direction of the away-side 
jet~\cite{Chatrchyan:2011sx}. 

Figure~\ref{aj-fig} (right) represents the distribution over the leading jet energy $E_T^{j1}$. This distribution seems quite stable. It does not depend significantly on jet quenching scenario, and is similar for PYQUEN, PYTHIA and CMS data. It confirms that the considered dijet configurations are dominated by a ``surface emission'', when the dijet production vertex is close to the surface of the nuclear overlapping area. In this case one partonic jet escapes from the dense zone almost without interactions and produces the leading hadronic jet, while a second partonic jet is affected by medium-induced multiple interactions and produces the ``quenched'' hadronic jet. 

Note that the absolute number of dijet events (having leading jet with $E_T^{j1}>$120 GeV and away-side jet with 
$E_T^{j2}>$50 GeV) is suppressed in PYQUEN as compared with PYTHIA due to reducing the number of away-side jets 
observed above the threshold 50 GeV (it corresponds to increasing the ``monojet'' events). The dijet suppression factor depends 
on the radiation angular spectrum and is estimated as 0.53, 0.36 and 0.29 for PYQUEN options 
(\ref{sar}), (\ref{war}) and (\ref{x-war}) respectively. 

\section{Conclusions}

The first LHC data on high-p$_T$ hadron and dijet spectra in PbPb collisions at $\sqrt s_{\rm NN}=2.76$ TeV
are analyzed in the frameworks of PYQUEN jet 
quenching model. The presented studies for the nuclear modification factor of high-p$_T$ hadrons and 
the imbalance in dijet transverse energy support the supposition that the intensive wide-angular (``out-of-cone'') 
medium-induced partonic energy loss is seen in central PbPb collisions at the LHC. This interpretation is 
in the agreement with the recent observation of CMS collaboration that the dijet momentum balance is 
recovered when integrating low transverse momentum particles distributed over a wide angular range 
relative to the direction of the away-side jet. 

However, there are still some complications for a more quantitative interpretation of the jet quenching pattern at the LHC. The measured by ALICE nuclear modification factor $R_{\rm AA}$ is affected by large systematic uncertainties due to the absence of pp data at the same energy. Moreover, the contribution of nuclear shadowing to $R_{\rm AA}$ seems visible at least up to $p_T \sim$ 25 GeV/$c$ and should be taken into account. Concerning ATLAS and CMS data on the dijet asymmetry, simple generator level studies are likely not enough to quantify this observable accurately (jet finding details and high multiplicity background fluctuations may be important). 

\begin{acknowledgement}
Discussions with V.L.~Korotkikh, L.V.~Malinina, L.I.~Sarycheva, S.V.~Petrushanko and B.~Wyslouch are gratefully acknowledged. 
This work was supported by Russian Ministry for Education and Science (contract 02.740.11.0244), Grant of President of Russian Federation (No 4142.2010.2), Russian Foundation for Basic Research (grant 10-02-93118) and Dynasty Foundation. 
\end{acknowledgement}

\end{document}